\documentclass[doublecol]{epl2}
\usepackage{epsfig}
\usepackage{bm}
\usepackage{graphics}
\usepackage{graphicx}
\usepackage{epsfig}
\usepackage{amssymb}
\usepackage{amsmath}
\usepackage{amsfonts}
\usepackage{color}
\usepackage{float}
\usepackage{bm}

\shorttitle{Black hole thermodynamics in the presence of a maximal length and minimum measurable in momentum}
\shortauthor{B. Hamil \etal}
\institute{
  \inst{1} D\'{e}partement de TC de SNV, Universit\'{e} Hassiba Benbouali, Chlef, Algeria.
\\
  \inst{2} Department of Physics, University of Hradec Kr\'alov\'e,
Rokitansk\'eho 62, 500 03 Hradec Kr\'alov\'e, Czechia. \\
  \inst{3} Department of Physics, Akdeniz University, Campus 07058 Antalya, Turkey.
}
\abstract{
In this work, incorporating the effect of the minimum measurable in momentum and maximal length, we studied thermodynamics property of Schwarzschild black hole and the Unruh effect. {\color{red} According to this scenario, we see that the black hole temperature cannot be smaller than a certain minimum value of $ T_{\min} $.  Moreover,  we find that black hole mass cannot be larger than a maximum mass value of $M_{\max }$. Considering these findings first we  compute the corrected Hawking temperature versus the mass and  examine its characteristic behavior.  Then, we derive the  black hole's entropy and heat capacity. We find that the black hole is stable when $\frac{M_{\max }}{\sqrt{3}}<M<M_{\max }$. Finally, we examined the modified Unruh effect. We find that the modified Unruh temperature explicitly depends on $\alpha$.}}
\begin{document}

\title{Black hole thermodynamics in the presence of a maximal length and
minimum measurable in momentum}
\author{B. Hamil\inst{1} \and B.C. L\" utf\"uo\u{g}lu\inst{2,3} }
\maketitle

\section{Introduction}

In general, incompatibility between Einstein's general relativity and
quantum field theory is regarded as the most fundamental motivation for the
development of quantum gravity. Different approaches which are used to form
the quantum theory of gravity, namely string theory \cite%
{Veneziano,Witten,Amati1,Konishi,Gross}, noncommutative geometry \cite%
{Capozziello}, and loop quantum gravity \cite{Garay}, put forward the
necessity of a fundamental lower limit length value. One way of defining
minimal length value is the generalization of the Heisenberg uncertainty
principle (HUP) by modifying the position and momentum operators. It is
shown that such a generalization is not unique. \cite{Kempf1}. For example,
in one of the generalized uncertainty principle (GUP) scenarios, the
commutation of the position and momentum operators end up with a term
depending on the momentum operator instead of a constant on the Hilbert
space \cite{Mangano,Kempf,Hinrichsen}. Many applications according to
various GUP scenarios have been investigated in \cite%
{Adler,Nozari,Nozari1,Nozari2,Nozari3,e1,Banerjee,e2,e3,e4,e5,e6,e7,e8,Duttaa,e9,e10,Ong,Nouicer,Pu,e11,e12,e13}%
. In particular, in \cite{Pu}, the authors explored the influences of the
GUP scenario on the thermodynamics of black holes (BHs). They found that in
that scenario the Hawking radiation can stop, so that, remnants of BHs
occur. Therefore, the GUP scenarios are considered as appropriate scenarios
to resolve the information paradox problem of BHs.

On the other hand, quantum gravity effects are assumed to act also on the
large-scale dynamics of the universe. Phenomenologically, the effects of
quantum gravity at large distances can be encoded in another kind of
generalization scenario, namely in the extended uncertainty principle (EUP)
scenario, where the commutation relations of the position and momentum
operators yield to a position operator instead of a constant or a momentum
operator that appear in the HUP and GUP scenarios, respectively \cite{Achim}. In the EUP scenario, a minimal measurable momentum value emerges {%
\color{red} which corresponds} to an infra-red cutoff \cite{Achim,
Haye,Brett,Park,Mignemi,Etemadi,Behnaz}. Mignemi proved that the EUP
scenario can also be extracted from the definition of quantum mechanics on
an (A)dS background with an appropriate choice of the parameterization \cite%
{Mignemi}. Recently, Perivolaropoulos proposed a new generalized uncertainty
principle to all orders in the Hubble parameter \cite{Perivolaropoulos,
Skara}. {\color{red}That} scenario implies the presence of a minimum
measurable momentum and a maximum measurable length together. {\color{blue} The existence of a maximum observable measurable length limit, in other words an infrared cutoff,  emerges naturally in the context of cosmological particle's horizon  \cite{Duttaa, Faraoni2011, Davis2003}  or cosmic topology \cite{Luminet2016}. Note that  these effects are expected to be effective in the early Universe.}

In this paper we consider a maximal length and a minimal observable momentum
scenario with the modified Heisenberg algebra and {\color{red} investigate}
the effects of our choice on the thermodynamics of Schwarzschild black hole.
We construct the paper as follows: {\color{red} At first,} we introduce the
modified algebra. {\color{red} Next, we obtain} the thermodynamic functions
in the modified algebra. {\color{red} Then}, we examine the Unruh
temperature and conclude the article with a brief conclusion.

\section{Quantum mechanics in the presence of a maximal length and minimum
measurable in momentum}

In this manuscript we {\color{red} take} the modified commutation relation {%
\color{red} that} is introduced in \cite{Perivolaropoulos,Skara} {\color{red}
into account}.
\begin{equation}
\left[ X,P\right] =\frac{i\hbar }{1-\alpha X^{2}}.  \label{1}
\end{equation}%
Here, three parameters, namely the Hubble parameter, $H_{0}$, speed of
light, $c$, and a dimensionless parameter, $\alpha _{0}$, are used to define
$\alpha $ parameter {\color{red}in the form of :} $\alpha =\left( \frac{%
\alpha _{0}H_{0}}{c}\right) ^{2}$. In order to satisfy Eq. \eqref{1}, we
consider the following position and momentum operators
\begin{equation}
X=x,\text{ \ }P=\frac{\hbar }{i}\frac{1}{1-\alpha x^{2}}\frac{d}{dx}.
\end{equation}%
In the position space representation, {\color{red} with the help of the
properties $\left\langle X^{2n}\right\rangle \geqslant \left\langle
X^{2}\right\rangle ^{n}$ {\color {blue}(with $n> 0$)}, }it is
straightforward to construct the GUP as follows: {\small
\begin{eqnarray}
\left( \Delta X\right) \left( \Delta P\right) &\geqslant &\frac{\hbar }{2}%
\left\langle \frac{1}{1-\alpha X^{2}}\right\rangle ,  \notag \\
&\geqslant &\frac{\hbar }{2}\left( 1+\alpha \left\langle X^{2}\right\rangle
+\alpha ^{2}\left\langle X^{4}\right\rangle +\alpha ^{3}\left\langle
X^{6}\right\rangle +...\right) ,  \notag \\
&\geqslant &\frac{\hbar }{2}\bigg(1+\alpha \left[ \left( \Delta X\right)
^{2}+\left\langle X\right\rangle ^{2}\right]  \notag \\
&+&\alpha ^{2}\left[ \left( \Delta X\right) ^{2}+\left\langle X\right\rangle
^{2}\right] ^{2}  \notag \\
&+&\alpha ^{3}\left[ \left( \Delta X\right) ^{2}+\left\langle X\right\rangle
^{2}\right] ^{3}+...\bigg),  \notag \\
&\geqslant &\frac{\frac{\hbar }{2}}{1-\alpha \left[ \left( \Delta X\right)
^{2}+\left\langle X\right\rangle ^{2}\right] }.
\end{eqnarray}%
} In order to determine the minimum measurable momentum of this deformed
algebra, we take only the physical states into account which satisfy $%
\left\langle X\right\rangle =0$ condition. Then, we solve the reduced GUP
\begin{equation}
\left( \Delta X\right) \left( \Delta P\right) =\frac{\hbar }{2}\frac{1}{%
1-\alpha \left( \Delta X\right) ^{2}},  \label{mm}
\end{equation}%
for $\left( \Delta P\right) .$ {\color{red} We obtain} the following minimum
observable momentum value
\begin{equation}
\left( \Delta P\right) _{\min }=\frac{3\sqrt{3}}{4}\hbar \sqrt{\alpha },
\end{equation}%
{\color{red} and} hence, the maximum measurable length value:%
\begin{equation}
\left( \Delta X\right) _{\max }=\ell _{\max }=\frac{1}{\sqrt{\alpha }}.
\end{equation}%
In this deformed scenario, the usual completeness and inner product
definitions between two states change with the following ones:
\begin{eqnarray}
1 &=&\int_{-\ell _{\max }}^{\ell _{\max }}\left( 1-\alpha x^{2}\right)
\left\vert x\right\rangle \left\langle x\right\vert dx, \\
\left\langle \psi \right. \left\vert \varphi \right\rangle &=&\int_{-\ell
_{\max }}^{\ell _{\max }}dx\left( 1-\alpha x^{2}\right) \psi ^{\ast }\left(
x\right) \varphi \left( x\right) .
\end{eqnarray}%
Here, the weight function, $\left( 1-\alpha x^{2}\right) $, is required for
the symmetry of the operators $X$ and $P$.

\section{Black holes}

\begin{figure}[tbph]
\centering
\includegraphics[scale=0.4]{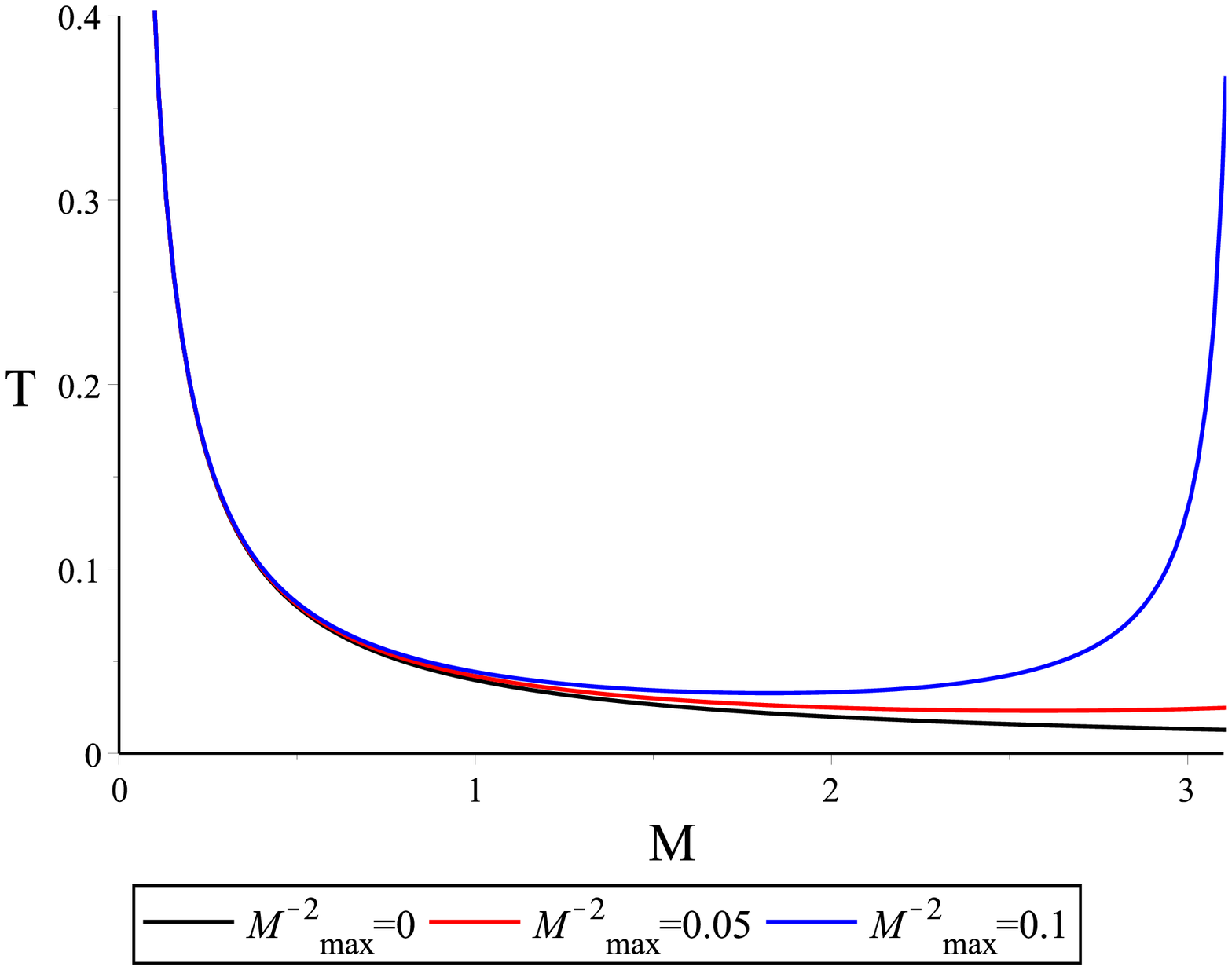}
\caption{Temperature-mass function for $\hbar =c=G=K_{B}=1$.}
\label{Fig1}
\end{figure}

In this section we examine the thermodynamics of a Schwarzschild BH under
the deformed scenario that is described above. In so doing, we consider the
following metric.
\begin{equation}
ds^{2}=-\left( 1-\frac{2MG}{rc^{2}}\right) c^{2}dt^{2}+\left( 1-\frac{2MG}{%
rc^{2}}\right) ^{-1}dr^{2}+r^{2}d\Omega ^{2}.
\end{equation}%
Here, $M$ denotes the mass of BH, $\Omega $ represents the solid angle, and $%
G$ is the Newton universal gravitational constant. {\color{red} From} this
line element, the event horizon can be written as%
\begin{equation}
r_{S}=\frac{2MG}{c^{2}}.  \label{max}
\end{equation}%
According to near-horizon geometry considerations one can set $\left( \Delta
X\right) =2\pi r_{S}$, thus Eq. \eqref{max} leads to a maximum horizon
radius and a mass value for the BH in the form of {\color{blue} \cite{Perivolaropoulos}}%
\begin{equation}
\left( r_{S}\right) _{\max }=2\pi \ell _{\max }\simeq 10^{26}\text{\textrm{m}%
};\text{ \ \ }M_{\max }=\frac{c^{2}\ell _{\max }}{4\pi G}\simeq 10^{52}\text{%
\textrm{kg}}.
\end{equation}%
{\color{blue}Note that the particle horizon is correlated with the length scale of the boundary between the unobservable and the observable regions of the Universe \cite{Perivolaropoulos}.} Then, we employ the temperature expression of any massless quantum particle
near the Schwarzschild BH horizon
\begin{equation}
T=\frac{c}{K_{B}}\left( \Delta P\right) ,  \label{tem}
\end{equation}%
to estimate a minimal observable temperature value of the BH via the minimum
uncertainty in momentum.
\begin{equation}
T_{\min }=\frac{c}{K_{B}}\left( \Delta P\right) _{\min }\simeq 10^{-29}\text{%
\textrm{K}}.
\end{equation}%
Next, we investigate the Hawking temperature of the BH. By substituting {%
\color{red} Eqs. \eqref{max} and \eqref{tem} into} Eq. \eqref{mm} we derive
the modified Hawking temperature in terms of the ordinary one, $T_{0}=\frac{%
\hbar c^{3}}{8\pi K_{B}MG}$, as follows:
\begin{equation}
T_{H}=\frac{T_{0}}{1-\frac{M^{2}}{M_{\max }^{2}}}.  \label{htem}
\end{equation}%
{\color{red} It is worth noting that} the term in the denominator modifies
the standard Hawking temperature. {\color{red} There, we observe a critical
mass value,}
\begin{equation}
M_{cr}=M_{\max }.
\end{equation}%
{\color{red} which plays a key role. If it has a finite value, it acts like
a cut-off because above this value the BH temperature becomes negative. On
the other hand, when it has an infinite value, $M_{\max }\rightarrow \infty $%
, then} the Hawking temperature reduces to the usual one \cite%
{Hawking,Bekenstein,35}. In the limit case where the mass term is close to {%
\color{red} the critical} mass value, the modified Hawking temperature
reaches very large values compared to usual one. If the mass term is very
small than {\color{red} the critical mass} value, then one can expand the
denominator term to
\begin{equation}
T_{H}\simeq T_{0}\left[ 1+\left( \frac{M}{M_{\max }}\right) ^{2}+\left(
\frac{M}{M_{\max }}\right) ^{4}+...\right] .  \label{htem1}
\end{equation}%
{\color{red}We present the variation of the modified Hawking temperature
versus the BH mass for different $M_{\max }^{-2}$ values in Fig. \ref{Fig1}.
We observe that} the temperature is divergent not only as $M\rightarrow 0$
but also as $M\rightarrow M_{\max }.$ {\color{red} In addition, we see the
presence of a minimum temperature value, $T_{\min }=\frac{3\sqrt{3}}{4}\frac{%
\hbar \sqrt{\alpha }c}{K_{B}}$. This value is achieved at $M=\frac{M_{\max }%
}{\sqrt{3}},$ and below that temperature value solution does not exist.}

\begin{figure}[tbph]
\centering
\includegraphics[scale=0.4]{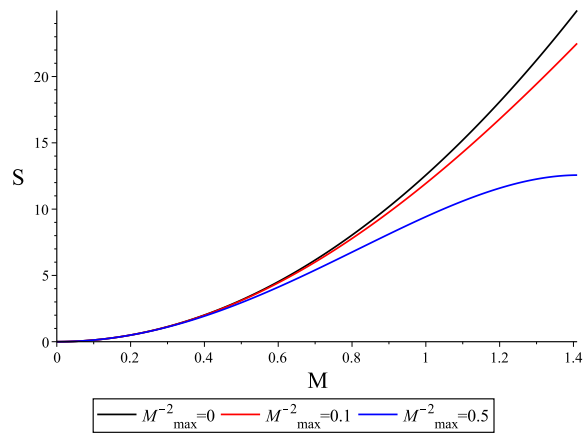}
\caption{BH's entropy versus mass for $\hbar =c=G=K_{B}=1$.}
\label{Fig2}
\end{figure}

Next, we determine the BH entropy from the first law of the BH
thermodynamics which is defined in the form of:%
\begin{equation}
S=c^{2}\int \frac{dM}{T}.  \label{entr}
\end{equation}%
After substituting Eq. (\ref{htem}) into Eq. (\ref{entr}) and performing the
integration, we obtain the GUP-corrected BH entropy as
\begin{equation}
\frac{S_{H}}{K_{B}}=\frac{S_{0}}{K_{B}}\left( 1-\frac{S_{0}}{2S_{\max }}%
\right) .  \label{hentr}
\end{equation}%
Here, $\frac{S_{0}}{K_{B}}=4\pi \frac{M^{2}}{M_{P}^{2}}$ is the
semi-classical Bekenstein-Hawking entropy for the Schwarzschild BH \cite%
{Hawking,Bekenstein,35} and $S_{\max }=4\pi \bigg(\frac{M_{\max }}{M_{P}}%
\bigg)^{2}$ is the maximum entropy value. We would like to remark that the
examined modification reproduces a correction term with a negative sign.
Then, we express the entropy (\ref{hentr}) in terms of the area of the
horizon, $A=4\pi r_{S}^{2}=4\ell _{P}\frac{S}{K_{B}}$. We find that Eq. (\ref%
{hentr}) can be described as%
\begin{equation}
\frac{S_{H}}{K_{B}}=\frac{A_{0}}{4\ell _{P}}\left( 1-\frac{A_{0}}{2A_{\max }}%
\right) ,  \label{area}
\end{equation}%
where $A_{\max }=4\ell _{P}\frac{S_{\max }}{K_{B}}$. We note that in the
limit of $\alpha \rightarrow 0$ it reduces to the famous area theorem. {%
\color{blue} In the present framework, we observe that a logarithmic-area correction term does not rise, while
in other approaches such as string theory, loop quantum gravity, effective models
with GUP \cite{Adler} and/or modified dispersion relations \cite{Mandanici} do. In addition, we find that the entropy gets its maximum value, $S_{H}=2\pi \left( \frac{M_{\max }}{M_{P}}\right) ^{2}$, when
the BH mass tends to $M_{\max }$}. To have a better knowledge of the characteristic behavior of the GUP-corrected BH entropy we plot the entropy,
Eq. (\ref{hentr}), versus the BH mass for different values of {\color{red} $%
M_{\max }^{-2}$} in Fig. \ref{Fig2}.
{\color{blue}Clear from the figure, in the presence of a maximal length and minimum measurable in momentum,  the entropy increases but with a decreasing rate lower than the standard case  and it gets its maximum when the BH mass goes to $ M_{max}$.}
Finally, we proceed to compute the modified heat capacity of the BH. To do
that, we employ the following relation
\begin{equation}
C=c^{2}\frac{dM}{dT}=\left( \frac{1}{c^{2}}\frac{dT}{dM}\right) ^{-1}.
\end{equation}%
After performing the simple algebra, We arrive at
\begin{equation}
C_{H}=C_{0}\frac{\left( 1-\left( \frac{M}{M_{\max }}\right) ^{2}\right) ^{2}%
}{1-3\left( \frac{M}{M_{\max }}\right) ^{2}},  \label{heat}
\end{equation}%
where $C_{0}=-\frac{8\pi K_{B}M^{2}}{M_{P}^{2}}$ is the standard expression
of the heat capacity. We observe that in the absence of modification, Eq. (%
\ref{heat}) reduces to the usual form of the BH heat capacity. In Fig. \ref%
{Fig3}, we demonstrate the behavior of the derived heat capacity function
versus mass for different values of $M_{\max }^{-2}$.
\begin{figure}[hbtp]
\centering
\includegraphics[scale=0.4]{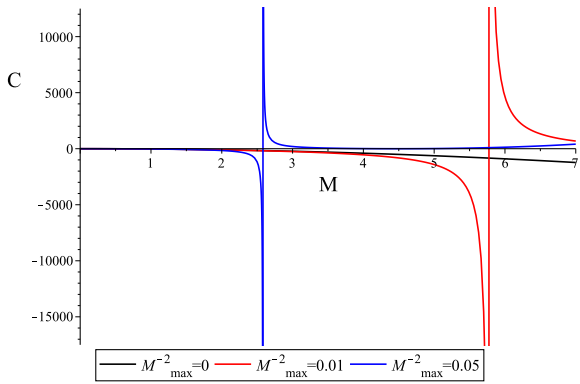}
\caption{BH's heat capacity versus mass for $\hbar =c=G=K_{B}=1$.}
\label{Fig3}
\end{figure}

In the interval of $0<M<\frac{M_{\max }}{\sqrt{3}}$, we see that the heat
capacity is negative and its value decreases faster with higher values of $%
\alpha $ and $M$. Contrarily for $M>\frac{M_{\max }}{\sqrt{3}}$ the heat
capacity is positive and reaches zero when $M\rightarrow M_{\max }.$ As it
is a well-known fact that if a black hole has a positive valued heat
capacity function, then it is assumed to be a stable. Alike, it is presumed
to be unstable when it has a negative valued specific heat. Therefore, we
conclude that BH can be unstable for $0\leqslant M\leqslant \frac{M_{\max }}{%
\sqrt{3}}$ and stable for $\frac{M_{\max }}{\sqrt{3}}\leqslant M\leqslant
M_{\max }$.

The collapse of a BH ends as soon as the heat capacity function tends to
zero. So that, the mass of the {\color{red}BH} remains the same. This mass
value is called the remnant mass, {\color{red}$M_{rem}$}. Its value can be
achieved by solving the following equation%
\begin{equation}
C_{H}=C_{0}\frac{\left( 1-\left( \frac{M}{M_{\max }}\right) ^{2}\right) ^{2}%
}{1-3\left( \frac{M}{M_{\max }}\right) ^{2}}=0,
\end{equation}%
In our case, the remnant mass is equal to the critical mass.
\begin{equation}
M_{rem}=M_{\max }.
\end{equation}%
{\color{blue}Before ending this section, it should be noted that the GUP scenario is characterised by the presence of a minimum value for the horizon radius and minimum mass $M_{\min}$. In addition, the heat capacity of the BH vanishes at the end point of the evaporation process which is characterized by BH remnant of mass $M_{\min}$ \cite{Adler}. In the present framework the BH can be stable only for masses around $M_{\max}$.}

\section{Unruh effect}

In this section, we examine the Unruh effect in this deformed scenario. We
start by substituting $\Delta P=\frac{\Delta E}{c}$ in Eq.(\ref{1}). We find
\begin{equation}
\left( \Delta E\right) =\frac{\hbar c}{2\left( \Delta X\right) }\frac{1}{%
1-\alpha \left( \Delta X\right) ^{2}}.
\end{equation}%
Then, we use the minimal distance $\left( \Delta X\right) $ along which each
particle must be accelerated to create $N$ particle \cite{e7}
\begin{equation}
\left( \Delta X\right) =\frac{2Nc^{2}}{a}.
\end{equation}%
After recalling the well-known relation, $\left( \Delta E\right) =\frac{3}{2}%
K_{B}T$, we find
\begin{equation}
T=T_{U}\phi \left( T_{U}\right) ,
\end{equation}%
where
\begin{equation}
\phi \left( T_{U},\alpha \right) =\frac{1}{1-\frac{\hbar ^{2}\alpha c^{2}}{%
9K_{B}^{2}T_{U}^{2}}},
\end{equation}
{\color{blue}and $T_{U}=\frac{\hbar a}{2\pi K_{B}}$ is the well-known Unruh
temperature,  while $a$ is the acceleration of the frame.  Finally, as indicated in \cite{son,Petruzziello}, the geometrical interpretation of quantum mechanics via a quantization model implies the existence of maximal acceleration which is conducted to a modification of the Heisenberg uncertainty principle. In a similar way, the present framework is characterised by presence of a minimum energy value
\begin{eqnarray}
\left( \Delta E\right) _{\min }=\frac{3}{2}K_{B}T_{\min }.
\end{eqnarray}%
Via this relation, we get a maximal bound value for the acceleration:
\begin{eqnarray}
a_{\max }\leq \frac{2\pi }{\hbar }K_{B}T_{\min }.
\end{eqnarray}%
This idea and analysis are in agreement with  \cite%
{Caia,Caianiello,Sharma}.}

\section{Conclusion}

In this manuscript we consider a generalized uncertainty principle out of
Heisenberg uncertainty principle that leads to a maximal length as well as a
minimum observable momentum value. Then, we examine the thermodynamic of a
Schwarzschild black hole. We find a minimal temperature and a maximum mass
value. We derive the modified Hawking temperature in terms of the usual one.
After demonstrating the mass-temperature function, we derive the entropy and
specific heat functions. We observe that both functions has two different
characteristic behaviors in two different interval. We find that the black
hole is first unstable until a critical mass value. After the mass exceeds
the critical mass value it becomes stable. In addition, we explore the
presence of a remnant mass value {\color{red} and find it at $M=M_{\max }$
value.} Finally we obtain the Unruh temperature in the deformed algebra. We
show that the modified Unruh temperature has a similar characteristic with
the usual one.

\end{document}